\documentclass[english,a4paper]{article}
\usepackage[T1]{fontenc}
\usepackage[cp1250]{inputenc}
\usepackage{amssymb}
\usepackage{hyperref}
\usepackage{amsfonts}
\usepackage{graphicx}

\begin{document}
\title{Exploring the Physical Limits of Saturation Contrast in Magnetic Resonance Imaging}

\author{M. Lapert, Y. Zhang, M. A. Janich, S. J. Glaser\footnote{Department of Chemistry, Technische Universit\"at
M\"unchen, Lichtenbergstrasse 4, D-85747 Garching, Germany}, D.
Sugny\footnote{Laboratoire Interdisciplinaire Carnot de Bourgogne
(ICB), UMR 5209 CNRS-Universit\'e de Bourgogne, 9 Av. A. Savary,
BP 47 870, F-21078 DIJON Cedex, FRANCE}}


\maketitle




\newpage

\begin{abstract}
Magnetic Resonance Imaging has become nowadays an indispensable
tool with applications ranging from medicine to material science.
However, so far the physical limits of the maximum achievable
experimental contrast were unknown. We introduce an approach based
on principles of optimal control theory to explore these physical
limits, providing a benchmark for numerically optimized robust
pulse sequences which can take into account experimental
imperfections. This approach is demonstrated experimentally using
a model system of two spatially separated liquids corresponding to
blood in its oxygenated and deoxygenated forms.
\end{abstract}

Since its discovery in the forties, Nuclear Magnetic Resonance
(NMR) has become a powerful tool \cite{ernst, spin} to study the
state of matter in a variety of domains extending from biology and
chemistry \cite{billion} to solid-state physics and quantum
computing \cite{chuang,vander}. The power of NMR techniques is
maybe best illustrated by medical imaging \cite{MRI}, where it is
possible e.g. to produce a three-dimensional picture of the human
brain. NMR spectroscopy and Magnetic Resonance Imaging (MRI)
involve the manipulation of nuclear spins via their interaction
with magnetic fields. All experiments in liquid phase can be
described in a first approach as follows. A sample is held in a
strong and uniform longitudinal magnetic field denoted $B_0$. The
magnetization of the sample is then manipulated by a particular
sequence of transverse radio-frequency magnetic pulses $B_1$ in
order to prepare the system in a particular state. The analysis of
the  radio-frequency signal that is subsequently emitted by the
nuclear spins leads to information about the structure of the
molecule and its spatial position. One deduces from this simple
description that the crucial point of this process is the initial
preparation of the sample, i.e.\ to design a corresponding pulse
sequence to reach this particular state with maximum efficiency.
The maximum achievable efficiency can be determined for the
transfer between well defined initial and target states
\cite{univ_bound} if relaxation effects can be neglected. In
imaging applications, where relaxation forms the basis for
contrast, a very large number of different strategies have been
proposed and implemented so far with the rapid improvement of NMR
and MRI technology \cite{spin, MRI}. However, there was no general
approach
 to provide the maximum possible performance and
the majority of these pulse sequences have been built on the basis
of intuitive and qualitative reasonings or on inversion methods
such as the Shinnar-Le Roux algorithm \cite{SLR}. Note that this
latter can be applied only in the case where there is no
relaxation effect and radio-frequency inhomogeneity.

A completely different point of view emerges if this
problem is approached from an optimal control perspective. Optimal control
theory was created in its modern version at the end of the 1950s with the Pontryagin Maximum Principle (PMP)
\cite{pont,bonnard,jurdjevic}. Developed originally for problems
in space mechanics, optimal control has become a key tool in a
large spectrum of applications including engineering,  biology and economics. Solving
an optimal control problem leads to the determination of a
particular trajectory, that is a solution of an associated
Hamiltonian system constructed from the PMP and satisfying given
boundary conditions. This approach has found remarkable
applications in quantum computing and NMR spectroscopy, but its application to MRI has been
limited to the numerical design of slice-selective 90$^\circ$ and 180$^\circ$ pulses \cite{encyclopedia}.

Despite the efficiency of MRI techniques currently used in clinics,
some aspects still pose fundamental problems of both theoretical and practical interest. The enhancement of contrast remains one of the
crucial questions for improving  image quality and
the corresponding medical diagnosis. The use of particular pulse
sequences to generate image contrast based on relaxation rates
is not new in MRI, since this
question was raised at the beginning of the development of MRI in
the 1970s. Different strategies have been proposed, such as
the Inversion Recovery sequence \cite{bydder,bydder2} for $T_1$ contrast and pulses for ultra short echo time experiments 
for $T_2$ contrast \cite{bydder3} (See Eq. (\ref{eq1}) for the
definition of $T_1$ and $T_2$ parameters). Here, we go beyond such
intuitive methods by using the powerful machinery of optimal
control, which provides in this case not just an improved
performance but an estimate of the global optimum, i.e. the best
possible contrast within the experimental constraints (see the
supplementary material for mathematical details). This optimized
contrast is demonstrated in a laboratory benchmark experiment.

In its simplest form, the contrast problem can be stated by
assuming that the signal is composed of two different
contributions. We consider as a benchmark example the case of
(a) oxygenated vs. (b) deoxygenated blood, where the spins that are probed are
the ones of the hydrogen atoms of water (H$_2$O).
This is e.g.  an important issue in functional studies of the human brain.
These spins have
different relaxation rates due to the interaction with other
molecules such as hemoglobin in its oxygenated or non-oxygenated
form, leading thus to two different signatures of the relaxation
dynamics of the magnetization, which is governed by the Bloch
equations:
\begin{equation}\label{eq1}
\left\{\begin{array}{rcl}
\frac{dM_x^i}{dt} & = &  -\omega M_y^i+ \omega_{y} M_z^i - M_x^i/T_2^i\\   
\frac{dM_y^i}{dt} & = &  \omega M_x^i- \omega_{x} M_z^i - M_y^i/T_2^i\\
\frac{dM_z^i}{dt} & = & - \omega_{y} M_x^i  +  \omega_{x} M_y^i  -
(M_z^i - M_0)/T_1^i,
\end{array} \right.
\end{equation}
where $\vec{M}^i=(M_x^i,M_y^i,M_z^i)$ is the magnetization vector
considered with $i=(a,b)$, $M_0$ is the equilibrium magnetization,
$T_1^i$ and $T_2^i$ the longitudinal and transverse relaxation
rates, $\omega$ the resonance offset and $\omega_x$ and $\omega_y$
the components in a rotating frame of the transverse magnetic
field along the $x$- and $y$- directions. While different
definitions of contrast exist in the literature \cite{MRI}, here
we consider a particular case that we call the \emph{saturation
contrast}, where the objective of the control problem is to find
the pulse sequence which completely suppresses the contribution of
one of the two magnetization vectors while maximizing the modulus
of the other. If such a pulse module can be found, it can be used
in combination with a large number of possible host sequences for
imaging and spectroscopy (see Chapter 14 of Ref. [6] for details).

\noindent \textbf{Results}\\
In this section, we first analyze the ideal situation of a
homogeneous ensemble of spin 1/2 particles irradiated on
resonance, which is described by Eq. (\ref{eq1}) with $\omega=0$.
We introduce the normalized vectors $\vec{V}^i=\vec{M}^i/M_0$ with
coordinates $(X^i,Y^i,Z^i)$ to eliminate the equilibrium
magnetization $M_0$. Based on numerical computations (see also
\cite{sugny2,sugny3,bonnardnew} for analytical details), we
restrict, without loss of generality, the dynamics to a meridian
plane by assuming $\omega_y=0$. Using advanced techniques of
geometric optimal control theory \cite{bonnard,lapertglaser}, it
is possible to find the desired control field $\omega_x(t)$ which
provides the optimum contrast by a direct solution of the PMP. The
details of the theoretical approach are given in Section 1 of the
supplementary material.

\begin{figure}
\centering
\includegraphics[scale=0.4]{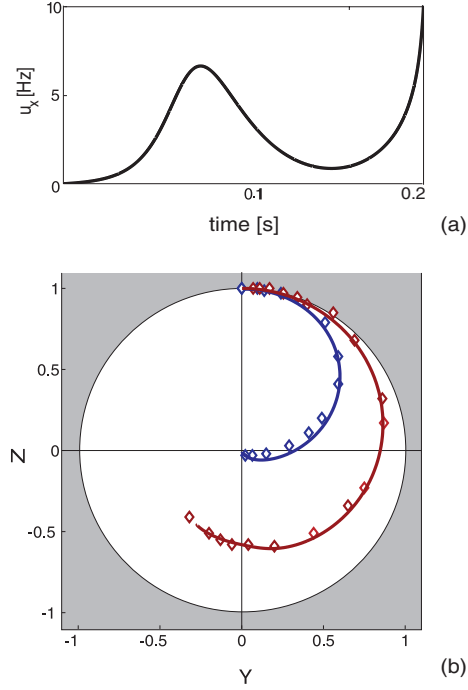}
\caption{(Color online) \textbf{Optimal pulse sequence and
trajectories for negligible $B_0$ and $B_1$ inhomogeneities.} {\bf
(a)}, The control amplitude $u_x(t)=\omega_x(t)/(2\pi)$ is shown
for the optimal pulse sequence to maximize $\vert \vec{V}^a \vert$
under the condition that $\vert \vec{V}^b\vert=0$. {\bf  (b)},
Corresponding simulated (solid curves) and experimental (open
diamonds) trajectories of $\vec{V}^a(t)$ (red)  and $\vec{V}^b(t)$
(blue) in the $(Y,Z)$ plane for a homogeneous ensemble of spin
particles.} \label{fig1}
\end{figure}

As a test case, we chose the typical relaxation parameters for (a)
oxygenated and (b) deoxygenated blood with identical $T_1$ values
($T^{a,b}_1=1.35$ s) but different $T_2$ values ($T^a_2=200$ ms,
$T^b_2=50$ ms). Note that in this situation, the conventional
Inversion Recovery experiment, which relies on $T_1$ differences
cannot provide useful contrast. However, applying the optimal
control approach  for these parameters the maximum of the modulus
of $\vec{V}^a$ (under the condition that $\vert \vec{V}^b\vert=0$)
is found to be $\vert \vec{V}^a\vert=0.4663$, representing the
maximum achievable saturation contrast in this system. Note that a
similar computation can be done to saturate the spin (a) and
maximize $\vert \vec{V}^b\vert$, with a final result of $\vert
\vec{V}^b\vert=0.4731$. We underline that this saturation contrast
is optimal in the sense that it is the physical upper limit that
can be reached within the experimental constraints given here by
the values of the relaxation rates. The shape of the optimal pulse
is shown in Fig. \ref{fig1}a. For our demonstration experiments,
we did not use actual blood samples but prepared two different
solutions with similar physical characteristics (see the Methods
section).  As the experimental $T_2^{\ast}$ values \cite{spin} of
the test sample were only about 2/3 of the $T_2$ values assumed
for the optimized pulse, the pulse duration and amplitude were
scaled by 2/3 and 3/2, respectively, resulting in a scaled pulse
duration of 0.214 s and a maximum pulse amplitude of the order of
10 Hz (see Fig. \ref{fig1}a). Within the experimental accuracy, we
have checked that, in this case, this scaling procedure provides
the same contrast as the optimal solution. The optimal control
field  was implemented experimentally as a shaped pulse on a
standard Bruker Avance III 600 MHz spectrometer and the
experimental trajectories were measured using two different
samples in different test tubes in order to approach ideal
experimental conditions with negligible magnetic field
inhomogeneities. Under these  conditions, the dynamics is
described with high accuracy by the Bloch equations (\ref{eq1}).
The simulated and experimental trajectories of the two
magnetization vectors $\vec{V}^a(t)$ and $\vec{V}^b(t)$ are shown
in Fig. \ref{fig1}b and a good match is found between theory and
experiment.

So far, we assumed that there is no experimental imperfection due
to magnetic field inhomogeneities, which are however not
negligible in realistic imaging experiments and have therefore to
be taken into account. Nevertheless, it is important to point out
that the analytical PMP-based optimal solution in the absence of
experimental imperfections gives an estimate of a previously
unavailable physical upper limit of the maximum achievable
saturation contrast, the inhomogeneities having a detrimental
effect on the final result. This bound, which can be determined
for any set of relaxation parameters, is thus relevant for any
imaging contrast problem. Hence, it gives a fundamental benchmark
to assess the performance of standard methods and numerically
optimized pulse sequences in this domain.

\begin{figure}
\centering
\includegraphics[scale=0.15]{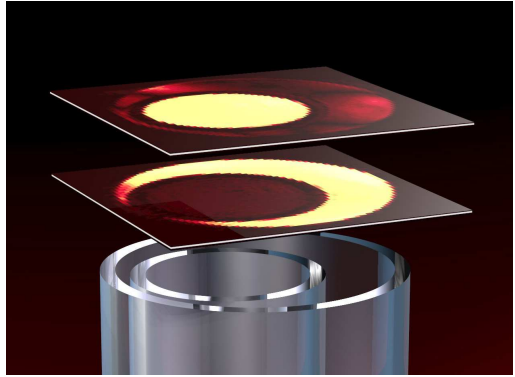}
\caption{(Color online) \textbf{Geometry of the test sample used
for the imaging experiments.} In the micro imaging experiments,
the sample consists of two test tubes with outer diameters of 5 mm
and 8 mm. The outer and inner volumes were respectively filled
with the two solutions corresponding to (a) oxygenated  and (b)
deoxygenated blood. The two slices represent the experimental
results after the saturation of the samples (a) (top) and (b)
(bottom). (see also the results of Fig. \ref{fig4})} \label{fig2}
\end{figure}

\begin{figure}
\centering
\includegraphics[scale=0.3]{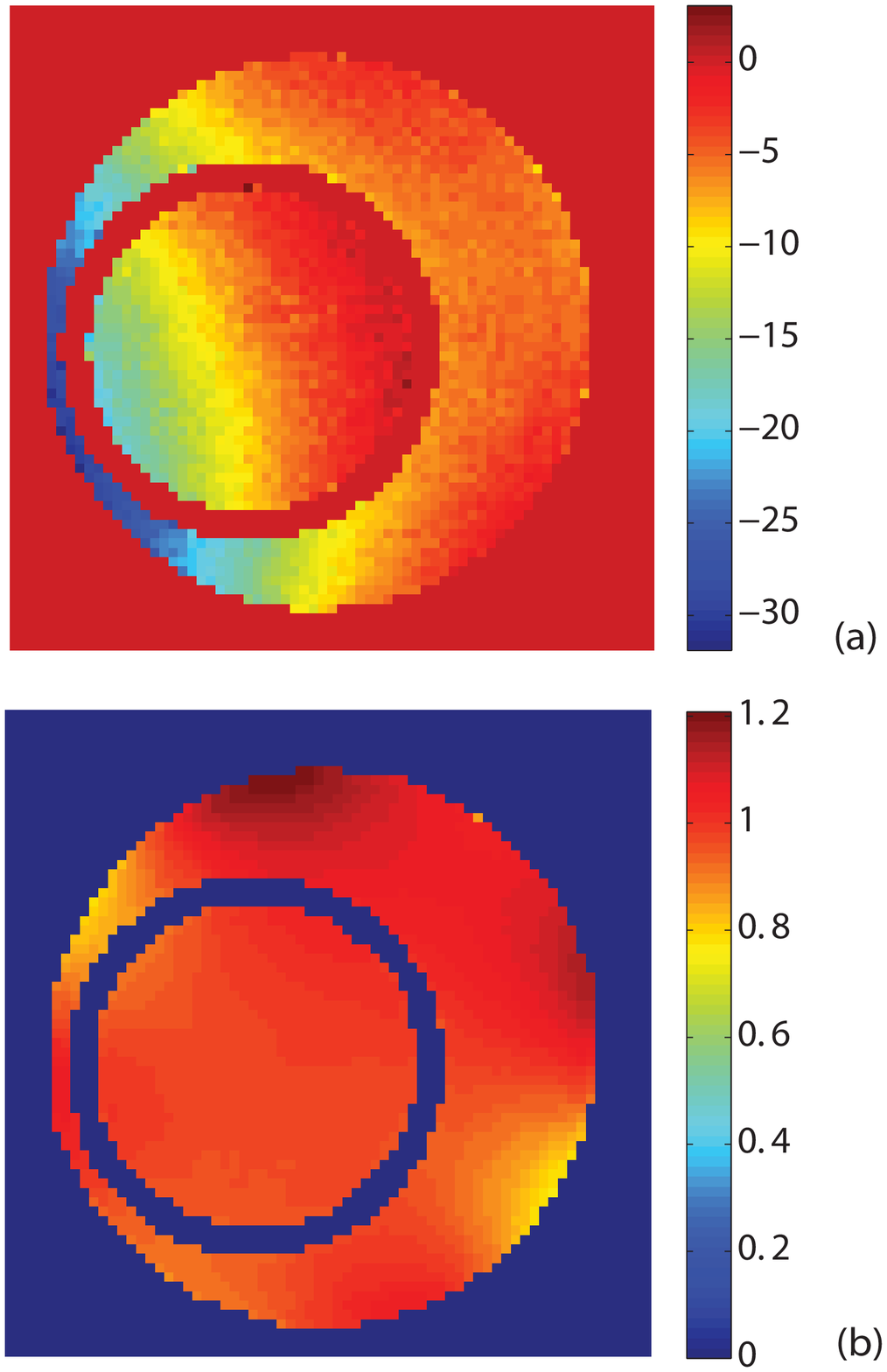}
\caption{(Color online) \textbf{Experimental spatial $B_0$ and
$B_1$ distributions}. Spatial distributions of the $B_0$ (a) and
$B_1$ (b) amplitudes in the central slice of the sample. The $B_0$
variation is represented by the corresponding $^1$H frequency
offsets in units of Hz as a function of the spatial position,
while the $B_1$ variation is described by a dimensionless scaling
factor.} \label{fig3}
\end{figure}

To perform a realistic imaging experiment, we designed a test
sample consisting of a small test tube in a larger tube with an
outer diameter of 8 mm, forming two compartments filled with
solutions (a) and (b) corresponding to the relaxation rates of
deoxygenated  and oxygenated blood, see Fig. \ref{fig2} for a
schematic representation. The experiments were performed using the
same spectrometer as described above, equipped with a
micro-imaging unit. The details are given in Section 2 of the
supplemental material. Figure \ref{fig3} shows the experimentally
measured spatial $B_0$ and $B_1$ distributions in the central
slice of the sample. The variation of $B_0$ corresponds to
resonance frequency shifts $\omega$ between 0 and -30 Hz, while
the experimental scaling of the $B_1$ field (which is proportional
to the control amplitude ) is $\pm 20\%$.
\begin{figure}
\centering
\includegraphics[scale=0.4]{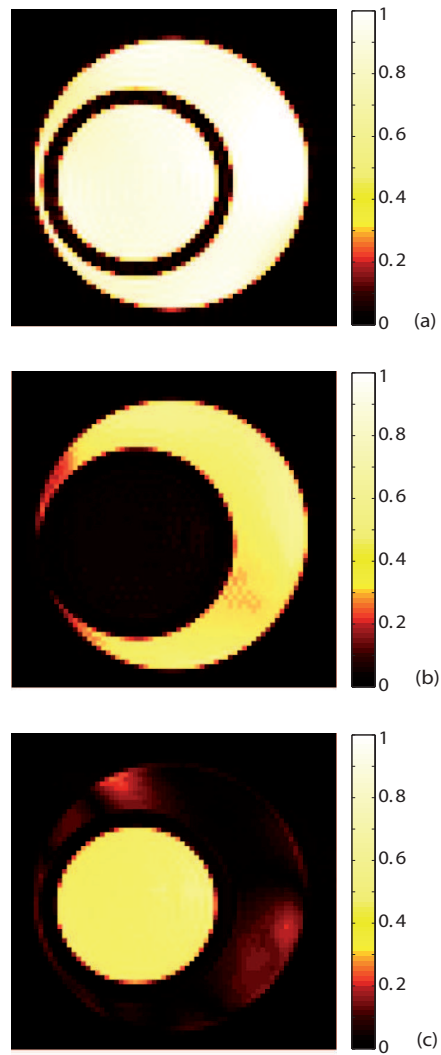}
\caption{(Color online) \textbf{Experimental implementation of the
robust optimal pulse}. Optimization of the contrast when the
deoxygenated (b) or the oxygenated (c) blood is saturated. Figure
(a) displays the reference image after the application of a 90
degree pulse.}\label{fig4}
\end{figure}

In general, the frequency offsets created by the $B_0$
inhomogeneities are negligible if they are dominated by the
amplitude of the control field (in units of Hz). However, the
control amplitude of the analytically optimal pulse sequence shown
in Fig. \ref{fig1}a is less than 10 Hz, which is smaller than the
resonance offset variation due to $B_0$ inhomogeneities (see Fig.
\ref{fig3}a) and therefore, the optimal sequence derived
analytically for an ideal case is not expected to work in the
experimental micro imaging setting. In addition, the experimental
variation of $B_1$ scaling was not considered in the analytical
solution and is also expected to have detrimental effects on the
pulse performance.

In order to take into account  the experimentally measured $B_0$
and $B_1$ distributions, numerically optimized pulse sequences
were computed with the GRAPE algorithm \cite{grape}, which is a
standard iterative algorithm to solve the optimization equations.
The pulses are designed to work for an ensemble of spins
approximately within the range of the $B_0$ and $B_1$
inhomogeneities experimentally measured. In the numerical
optimizations, we fixed the pulse duration to the duration of the
corresponding analytical pulse. Compared to the fundamental
contrast benchmark ($\vert V^a \vert= 0.47$ and $\vert V^b \vert=
0$) provided by the PMP-based analytical solution, the minimum
value (worst case) of $\vert V^a \vert$ for the considered range
of $B_0$ and $B_1$ inhomogeneities is $0.37$ (that is 79\% of the
physically maximum saturation contrast achievable) while the
maximum value (worst case) of the incompletely suppressed $\vert
V^b \vert$ is $0.054$. Conversely, when the goal is to saturate
spin (a) and to maximize the magnetization vector of spin (b), the
optimal pulse sequence yields $\vert \vec{V}^a\vert =0.047$ and
$\vert \vec{V}^b\vert =0.33$. Figure \ref{fig4} shows  resulting
experimental images which are in good agreement with simulated
data (see the supplementary material for the details of the
computation).

\noindent \textbf{Discussion}\\
We demonstrated to which extend saturation contrast based on
different relaxation times $T_1$ and $T_2$ can be maximized in
magnetic resonance imaging within given experimental constraints.
Starting from the analytic optimal solution of the homogeneous
case, we have then designed a particular pulse sequence using
numerical tools of optimal control theory to approach in a
realistic experiment this physical limit.
We emphasize that one of the main advantages of this
contrast enhancement is its general character since the optimal
control fields can be computed with standard routines published in
the literature and implemented on a standard NMR spectrometer without
requiring specific materials and process techniques.

The efficiency of this approach was shown in a laboratory
experiment using a model system for the relaxation parameters of
deoxygenated and oxygenated blood. The presented method fully
exploits the differences of both $T_1$ and $T_2$ to create the
maximum possible saturation contrast as opposed to conventional
approaches based on
 $T_1$ or $T_2$ differences.
The combined analytical and numerical optimal control approach is
not limited to the definition of saturation contrast (motivated by
typical magnitude mode imaging experiments) used here for
demonstration, but can also be applied to more general definitions
of relaxation-based contrast, e.g. for phase-sensitive images and
for a wide variety of possible host imaging sequences that can be
applied after the contrast pulse module \cite{MRI}. Furthermore,
the flexibility of the optimal control approach makes it possible
to include experimental constraints such as bounds on the control
amplitude or pulse energy or non-linear effects such as radiation
damping \cite{yun}.

We expect that the presented principles will find practical applications in MRI and in particular in medical imaging, where increased contrast and
sensitivity could not only help in the diagnosis but
could also reduce the required concentration of commonly used contrast
agents, which could be beneficial to the
patient.

\noindent \textbf{Methods}\\
\textbf{Experimental sample.} The relaxation properties of
oxygenated blood were approached by solution (a) consisting of
90\% D$_2$O, 10\% glycerol and  doped with CuSO$_4$ with
relaxation times $T_1^a=1.8$ s, $T_2^a=260$ ms (as determined from
CPMG-experiment) and $T_2^{*,a}=100$ ms (as determined from the
experimental line width). Deoxygenated blood was modeled by
solution (b) consisting  of 70\% D$_2$O, 30\% glycerol and doped
with CuSO$_4$ with $T_1^b=1.4$ s, $T_2^b=60$ ms and $T_2^{*,b}$=30
ms.\\
\textbf{Measurement method of the $B_0$ and $B_1$ field maps.}
Mapping of local $B_0$ offsets was accomplished by evaluating the
signal phase evolution between two echoes acquired in a dual-echo
gradient pulse sequence \cite{MRI}. The echo times of the 3D image
acquisition were $TE_1=1.5$ ms and $TE_2=11.5$ ms. Figure 3a of
the main text shows the $B_0$ field map in the central axial
slice. The amplitude of the $B_1$ excitation field was measured by
using the cosine-like dependence of the remaining signal after a
saturation pulse and fitting to a curve measured with multiple
saturation flip angles~\cite{Brunner}. The applied saturation flip
angles were $10^\circ, 20^\circ,  \dots , 300^\circ$.
Robust $B_1$ mapping was achieved by fitting a signal model to the
acquired slice-selective gradient echo signal in a linear
least-squares sense.\\
\textbf{Application of the optimized pulse sequence.} For the
imaging experiment, the H$_{\rm 2}$O content was increased in the
sample in order to have a sufficient signal-to-noise ratio. The
outer and inner volumes of the sample were filled with the
following solutions: "oxy sample II" (80\% D$_{\rm 2}$O, 10\%
H$_{\rm 2}$O, 10\% glycerol doped with CuSO$_{\rm 4}$ with
relaxation times of $T_1$=2.6 s and $T_2^*$=100 ms) and "deoxy
sample II" (60\% D$_{\rm 2}$O, 10\% H$_{\rm 2}$O, 30\% glycerol
doped with CuSO$_4$ with relaxation times of $T_1$=1.4 s and
$T_2^*$=30 ms). The optimal pulses were implemented into a
gradient echo pulse sequence \cite{gradientecho} without slice
selection. The experiments were performed with a 600MHz Bruker
Avance III spectrometer equipped with a micro-imaging unit. The
field of view (FOV) was 15mm x15 mm, and in the third dimension it
was limited to approximately 10 mm by coil sensitivity. The
repetition time (TR) is 7 s and the echo time (TE) is 2 ms with
the matrix size 128x128.  Figures S1 and S2 of the supplementary
material show simulated (d) and experimental (e) images using the
optimized pulse to saturate the "deoxy sample II" in the inner
cylinder (Fig. S1) and for the optimized pulse to saturate the
"oxy sample II" in the outer cylinder (Fig. S2).
%
%

\bibliography{biblio}

\vspace{1cm}

\noindent \textbf{Acknowledgment}\\
We are grateful to B. Bonnard and H. R. Jauslin for discussions and Franz Schilling for help in the imaging experiments.
S.J.G. acknowledges support from
the DFG (GI 203/6-1), SFB 631. S. J. G. thanks the Fonds der Chemischen Industrie. Experiments
were performed at the Bavarian NMR center at TU M\"unchen.\\
\noindent \textbf{Author contributions}\\
All authors contributed to the design and interpretation of the
presented work. Numerical computations have been done by M. L. and
Y.
Z., the construction of the experiment and to the acquisition of the data was performed by Y. Z., M. A. J. and S. J. G.\\
\noindent \textbf{Additional information}\\
Supplementary information accompanies this paper.\\
Competing financial interests: The authors declare no competing
financial interests.\\
\noindent \textbf{Author information}\\
Correspondence and requests for materials should be addressed to
D. S. (email: dominique.sugny@u-bourgogne.fr) and S. J. G. (email:
glaser@tum.de).
\end{document}